\documentstyle[11pt,fullpage,doublespace,epsf]{article}
 \DeclareMathSizes{11}{19}{13}{9}   

\makeatletter
\@addtoreset{equation}{section}
\makeatother

\begin{document}

   \title{ Riemann Hypothesis, Matrix/Gravity Correspondence and FZZT Brane Partition Functions}

\author{Michael McGuigan\\Brookhaven National Laboratory\\Upton NY 11973\\mcguigan@bnl.gov}
\date{}
\maketitle

\begin{abstract}
We investigate the physical interpretation of the Riemann zeta function as a FZZT brane partition function associated with a matrix/gravity correspondence. The Hilbert-Polya operator in this interpretation is the master matrix of the large $N$ matrix model. Using a related function $\Xi(z)$ we develop an analog between this function and the Airy function $Ai(z)$ of the Gaussian matrix model. The analogy gives an intuitive physical reason why the zeros lie on a critical line. Using a Fourier transform of the $\Xi(z)$ function we identify a Kontsevich integrand. Generalizing this integrand to $n \times n$ matrices we develop a Kontsevich matrix model which describes $n$ FZZT branes. The Kontsevich model associated with the $\Xi(z)$ function is given by a superposition of Liouville type matrix models that have been used to describe matrix model instantons. 
\end{abstract}

\section{Introduction}

It is an old idea that if a Hermitian operator can be found which has eigenvalues of the form $\lambda_n = -i(\rho_n - \frac{1}{2})$ where $\rho_n$ are the nontrivial zeros of the zeta function then the Riemann hypothesis would be true. This would follow because the eigenvalues of a Hermitian operator are real. No such operator has yet been found however.

In the Heisenberg matrix formulation of quantum mechanics \cite{Heisenberg} one represents observables with infinite matrices which are Hermitian. The eigenvalues of the matrix are what are measured in an experiment and hence are real. The infinite matrix can be constructed by forming an $N \times N$ matrix and taking the large $N$ limit. No such large $N$ matrix whose eigenvalues are related to the Riemann zeros has been found. 

One can also consider theories called  matrix models in which the dynamical variables are such large $N$ matrices \cite{Wigner}\cite{Dyson:1972tm}\cite{Martinec:2004td}. There is  a  remarkable correspondence between such matrix theories and continuous theories describing a quantum theory of world sheet gravity and low dimensional string theory. In such a correspondence invariants of the matrix theory  are related to  geometric observables in the world-sheet gravity usually through an integral transform \cite{Martinec:2004td}. This integral transform takes one from a variable in the expansion of the characteristic polynomial of the large $N$ Hermitian matrix description to a Liouville variable describing the size of the string or of a of a macroscopic loop of 2d gravity in the continuum description. The correspondence arises because Feynman graphs in the  matrix theory description can yield discrete representation of surfaces which become continuous as one takes $N$ to infinity \cite{'tHooft:1973jz}. 

In this paper we interpret the Riemann zeta function as being related to a particular observable in the matrix/gravity correspondence namely the FZZT brane partition function of a matrix model and interpret it's master matrix as the Riemann operator. The potential for this matrix model is more complicated than most of the ones considered in the literature. Nevertheless the techniques of simpler matrix theories can be applied to this case as well. This paper is organized as follows. In section 2 we discuss the master matrix approach to matrix models. We discuss some of the conceptual advantages of the approach as well as the difficulties. In section 3 we discuss introduce the FZZT brane partition function from the matrix model point of view. In section 4 we determine the Kontsevich integrand associated with the Riemann zeta function and develop an analogy between the Riemann zeta function and the Airy function which is the FZZT partition function of the $(2,1)$ minimal matrix model. In section 5 we discuss how to approximate the matrix model associated with Riemann zeta function using the generalized $(p,1)$ matrix model for large $p$ whose FZZT partition function is a generalized Airy integral. In section 6 we review the main conclusions of the paper.
 
\section{Master matrix}

If one can find a special  infinite Hermitian matrix $M_0$ such that:
\[
\Xi (z ) = \det (M_0  - z I)  
\]
where
\[
\Xi (z ) = \zeta (i z  + \frac{1}{2})\Gamma (\frac{{z }}{2} + \frac{1}{4})\pi ^{ - 1/4} \pi ^{ - iz /2} (
- \frac{{z ^2 }}{2} - \frac{1}{8})
\]
then the Riemann hypothesis would be true. This is because this function can be written in product form as:
\[
\Xi (z ) = \frac{1}{2}\prod\limits_n {(1 - \frac{{iz  + 1/2}}{{\rho _n }}} )
\]
The eigenvalues of the Hermitian matrix $M_0$ are denoted by $\lambda_n$ and are related to the Riemann zeros via $\rho_n = i\lambda_n + 1/2$. Then the product becomes:
\[
\Xi (z ) = \frac{1}{2}\prod\limits_n {(1 - \frac{{iz  + 1/2}}{{i\lambda _n  + 1/2}}} ) = \frac{1}{2}\prod\limits_n {\frac{{\lambda _n  - z }}{{\lambda _n  - i/2}}}
\]
This vanishes at the values $\lambda_n$ just as the formal determinant expression. The $\lambda_n$ are real if the matrix $M_0$ is Hermitian and thus the Riemann Hypothesis would be true. Unfortunately just as for the Riemann operator referred to above no such infinite matrix $M_0$ has ever been constructed.

The difficulty in constructing $M_0$ is somewhat similar to the difficulty in constructing a master field or master matrix in large $N$ field theory of matrix theory \cite{Gopakumar:1994iq}\cite{Gopakumar:1995bk}. A master field or master matrix is a special large $N$ matrix such that statistical averages of an observable can be computed by simply evaluating the observable on the the special large $N$ matrix. The reason that a master matrix exists is because at large $N$ expectation values factorize as:
\[
\begin{array}{l}
 \left\langle {O_1 O_2 } \right\rangle  = \left\langle {O_1 } \right\rangle \left\langle {O_2 } \right\rangle  + O(1/N^2 ) \\
 \left\langle {(O - \left\langle O \right\rangle )^2 } \right\rangle  = \left\langle {O^2 } \right\rangle  - \left\langle O \right\rangle ^2  = O(1/N^2 ) \\
 \end{array}
\]
where:
\[
\left\langle {O } \right\rangle  = \int DM O(M) e^{-V(M)}
\]
and $V(M)$ is a matrix potential. Thus  variances vanish so the observable's value is localized on a particular matrix as $N \to \infty$ just as
particle trajectories are localized on classical solutions as $\hbar$ goes to zero. Once such a master field is found the above observables are simply given by:
\[
\left\langle {O } \right\rangle  = O(M_0)
\]
There are several such observables in matrix theory. We discuss some of these in the next section.

For a general matrix model with potential $V(M)$ the master matrix can be written \cite{Gopakumar:1994iq}\cite{Gopakumar:1995bk}:
\[
M_0  = S^{ - 1} TS = S^{ - 1} (a + \sum\limits_{n = 0}^\infty  {t_n a^{ + n} )S}  
\]
where the similarity transformation $S$ is defined so that $M_0$ is Hermitian and the operators $a,a^+$ obey $[a,a^+] = I$. One can expand the master matrix as a function of the Hermitian operator $\hat x = a + a^{+}$ as:
\[
M_0 (\hat x) = g_1 \hat x + g_2 \hat x^2  +  \ldots
\]
One can also define an associated complex function:
\[
M_0 (y) = \frac{1}{y} + \sum\limits_{n = 0}^\infty  {t_n y^n }
\]
as well as a conjugate matrix $P_0$ that satisfies:
\[
[P_0 ,M_0 ] = I
\]
The Master matrix can be determined from the equation \cite{Gopakumar:1994iq}\cite{Gopakumar:1995bk}:
\[
(V'(M_0 (\hat x)) + 2P_0)\left| 0 \right\rangle  = 0
\]
Here $\left| 0 \right\rangle$ is the vacuum state annihilated by $a$. The master matrix is closely connected with the resolvent $R(z)$ and eigenvalue density $\rho(x)$ through:
\[
R(z) = Tr(\frac{1}{{z - M_0 }}) = \int {dx\frac{{\rho (x)}}{{z - x}}}  =  - \oint\limits_C {\frac{{dw}}{{2\pi i}}} \log (z - M_0 (w))
\]
The associated function $M_0(y)$ obeys the relation:
\[
R(M_0 (y)) = M_0 (R(y)) = y
\]
The function $yM_0(y)$ is the generating functional of connected Green functions for the generalized matrix model. While the concept of the master matrix is appealing, to construct the master matrix explicitly is equivalent to finding all the connected Green functions which amounts to solving the theory. This can be done for the potential $V(M) = Tr(M^2)$ but for the general matrix model is quite difficult.  In the next three sections we turn to other methods of dealing with the generalized matrix model  which are somewhat more tractable and apply them to the interpretation of the zeta function.

\section{FZZT brane}

One observable of matrix models is the exponentiated macroscopic loop or FZZT brane partition function \cite{Fateev:2000ik}\cite{Teschner:2000md}\cite{Giusto:2004mt}\cite{Ellwood:2005nt}. This is given by:
\[
B(z ) = \det (M - z I)
\]
This is the characteristic polynomial associated with the matrix $M$. It's argument $z$ can be complex. In the context of the Riemann zeta function $\zeta(s)$ the variable is related to the usual argument of the zeta function by $s = iz + \frac{1}{2}$.
Another observable is the macroscopic loop which is the transform of the Wheeler-DeWitt wave function defined on the gravity side of the correspondence \cite{Klebanov:2003wg}.
\[
W(z ) =  - Tr\log (M - z I) = \lim _{\varepsilon  \to 0}
(\int\limits_\varepsilon ^\infty  {\frac{{d\ell }}{\ell }} Tr(e^{\ell (-z I + M)} ) + \log \varepsilon )
\]
where $\epsilon$ is a UV cutoff.

The resolvent observable mentioned above is defined by:
\[
R(z ) = \frac{{\partial W(z )}}{{\partial z }} = Tr(\frac{1}{{M - z I}})
\]
Finally one has the inverse determinant observable defined in \cite{Klebanov:2003wg}. 

If a special master matrix $M_0$ can be found then expectation values such as
\[
\left\langle {B(z )} \right\rangle  = \left\langle {\left. {\det (M - z I)} \right\rangle } \right. = \int {DM\det (M - z I)e^{ - V(M)} } = \det (M_0  - z I) 
\]
reduce to evaluating the observable at $M_0$. In the context of the $\Xi(z)$ function the desired relation is of the form: 
\[
\Xi (z ) = \det (M_0  - z I) = \left\langle {B(z )} \right\rangle  = \left\langle {\left. {\det (M - z I)} \right\rangle } \right. = \int {DM\det (M - z I)e^{ - V(M)} }
\]

Some matrix potentials that have been considered are 
\[
V(M) = Tr(M^2)
\]
which describes 2d topological gravity or the (2,1) minimal string theory \cite{Witten:1989ig} \cite{Maldacena:2004sn} \cite{Kutasov:2004fg}. A quartic potential:
\[
V(M) = Tr( - M^2  + gM^4 )
\]
is used to describe minimal superstring string theory \cite{Seiberg:2004ei}\cite{Seiberg:2003nm}\cite{Fukuma:2006ny}\cite{Johnson:2003hy}. A more complicated matrix potential is
\[
V(M) =  - Tr(M + \log (I - M)) = \sum\limits_{m = 2}^\infty  {\frac{1}{m}Tr(M^m
)}
\]
which defines the Penner matrix model \cite{Penner}\cite{Distler:1990pg}\cite{Distler:1990mt}\cite{Imbimbo:1995ns}\cite{Matsuo:2005nw} and is used to compute the Euler characteristic of the moduli space of Riemann surfaces. 

Another matrix model that has been introduced is the Liouville matrix model \cite{Mukhi:2003sz}\cite{Imbimbo:1995yv} with potential given by:
\[
V(M)= Tr (\alpha M + \mu e^M)
\]
with cosmological constant $\mu$ so that:
\[
e^{-V(M)} = e^{-\alpha Tr M}e^{-\mu Tr e^M}
\]
In this paper we will encounter the matrix potential determined by:
\begin{equation}
e^{-U(M)} = \sum\limits_{q = 1}^\infty  {(q^4 \pi ^2 e^{2TrM}  - \frac{3}{2}} q^2 \pi e^{TrM})e^{ - q^2 \pi Tr( e^M) }
\end{equation}
The partition function for this matrix model can be seen as a superposition of partition functions of Liouville matrix models with cosmological constants of the form.
\[
\mu = q^2\pi
\]
for integer $q$. The origin of this particular matrix model and it's relation to the zeta function will be discussed in the next section.

\section{Kontsevich integrand}

To see how the matrix potential (3.1) arises it is helpful to consider how the coefficients of the characteristic polynomial observable $B(z)$ can be determined by expanding as a series in $z$. If the function $\Xi(z)$ is interpreted as a characteristic polynomial then one can obtain these coefficients from the expansion:
\[
\Xi (z )  = \sum\limits_{n = 0}^\infty  {a_{2n} \frac{{( - 1)^n }}{{\left(
{2n} \right)!}}} z ^{2n}
\]
where
\[
a_{2n}  = 4\int\limits_1^\infty  {d\ell (\ell ^{ - 1/4} f(\ell )(\frac{1}{2}\log \ell )^{2n} } )
\]
and 
\[
f(\ell ) = \sum\limits_{q = 1}^\infty  {(q^4 \pi ^2 \ell }  - \frac{3}{2}q^2 \pi )\ell ^{1/2} e^{ - q^2 \pi \ell }
\]
Inserting the coefficients $a_{2n}$ into $\Xi(z)$ and summing over $n$ we can represent $\Xi(z)$ as an integral transform:
\[
\Xi (z ) = 4\int\limits_1^\infty  {\frac{{d\ell }}{\ell }} \ell ^{(iz  + 1/2)/2} \sum\limits_{q = 1}^\infty  {(q^4 \pi ^2 \ell ^2  - \frac{3}{2}} q^2 \pi \ell )e^{ - q^2 \pi \ell }  = 4\int\limits_1^\infty  {\frac{{d\ell }}{\ell }}
\ell ^{(iz  + 1/2)/2} \ell ^{1/2} f(\ell )
\]
Defining the variable $\phi$ by $\ell = e^\phi$ we have:
\begin{equation}
\Xi[z] = \int {d\phi e^{iz\phi} } \sum\limits_{k = 1}^\infty  {(\pi ^2 k^4 }
e^{2\phi }  - \frac{3}{2}\pi k^2 e^{\phi } )e^{ - \pi k^2 e^\phi  }
\end{equation}
which is a well known integral expression for the function  $\Xi(z)$.

For the simple potential $V(M) = Tr(M^2)$ the exponentiated macroscopic loop observable (FZZT brane) can be computed. It is given by the Airy function \cite{Kutasov:2004fg}:
\begin{equation}
Ai(z) = \int DM det(M-zI) e^{-Tr(M^2)}= \int d\phi e^{iz\phi + i\phi^3\frac{1}{3}}
\end{equation}
Because this function is associated with an Hermitian matrix model it's zeros are real.
This is the analog of the Riemann hypothesis for $V(M) = Tr(M^2)$.
The similarity between the integral representations of (4.1) and (4.2) suggest an analogy between the Airy and zeta functions.

To illustrate a comparison between the Airy function and the zeta function consider figures 1 and 2. The zeros disappear as one moves off the critical line which corresponds to $z$ real in both cases. This suggests that the zeta function corresponds to a Hermitian matrix model. Table 1 illustrates the comparison on both sides of the correspondence. The question mark indicates the (substantial) missing information involved in a matrix/gravity approach to the Riemann hypothesis.

Qualitative differences exist between between the functions $Ai(z)$ and $\Xi(z)$. The $Ai(z)$ function is exponentially decaying to the positive $z$ axis. This is a result of Stokes phenomena where an exponentially growing form of the Airy function is completely absent in the right $z$ axis. For the $\Xi(z)$ function one does not see exponentially decaying function in the positive $z$ axis. Instead one has identical behavior in the positive and negative $z$ axis. One way to see the difference is to use a Riemann-Hilbert Problem approach to both functions \cite{Its}\cite{Konig}\cite{Brezin}\cite{Kitaev}\cite{Strahov}\cite{Gangardt}. In the case of the Airy function this leads to the differential equation \cite{Its}:
\[
Ai''(z) = zAi(z)
\]
whereas in the case of the $\Xi(z)$ function one does not obtain a differential equation but a discrete equation \cite{Its} :
\[
\Xi (z) = \Xi ( - z)
\]
Indeed it is known that the zeta function does not obey a finite order differential equation so this may be a possible explanation for the qualitative difference between the two functions. It would be interesting to explore further the differences between the two functions using the Riemann-Hilbert approach of and their interpretations as FZZT brane partition functions.
.
 
\begin{figure}[htbp]
  
   \centerline{\hbox{
   \epsfxsize=5.0in
   \epsffile{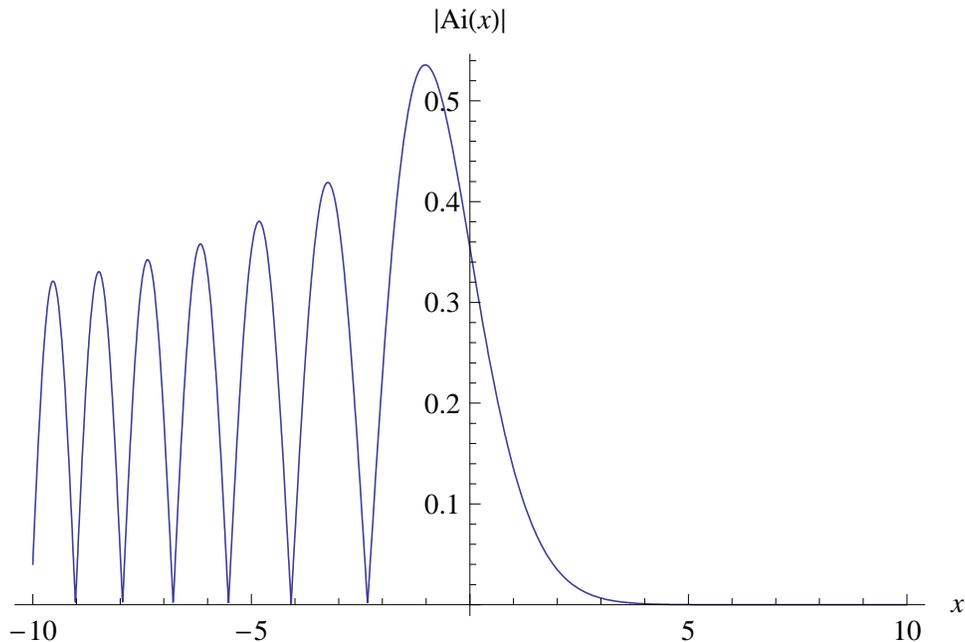}
     }
  }
 \caption{Magnitude of the Airy function on the real axis. The zeros are all located on the negative real axis. An intuitive way to understand this is that the Airy function is the FZZT brane partition function of a matrix model with potential $V=Tr(M^2)$ and Kontsevich integrand $e^{-U(\phi)} = \exp(i\phi^3/3)$}
             
  \label{fig1}
  
\end{figure}

\begin{figure}[htbp]
  
   \centerline{\hbox{
   \epsfxsize=5.0in
   \epsffile{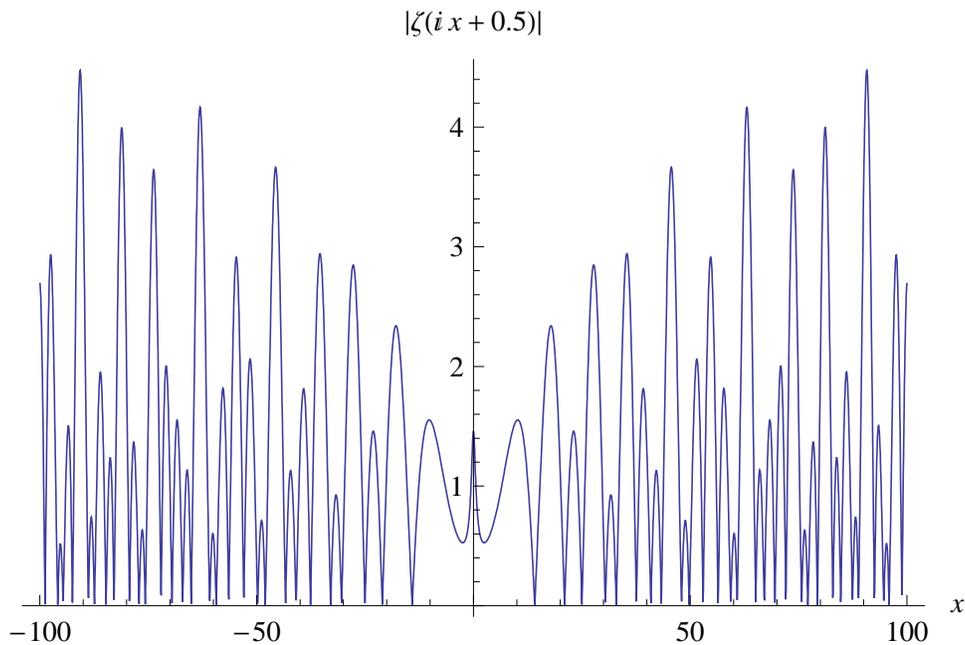}
     }
  }
 \caption{Magnitude of the function $\zeta(ix+1/2)$ on the real axis. The zeros are all symmetrically located on the real axis. An intuitive way to understand this is that the $\Xi$  function is the FZZT brane partition function of a matrix model with a suitably chosen potential $V(M)$ and Kontsevich integrand $e^{-U(\phi)}$.}
             
  \label{fig2}
  
\end{figure}

\begin{table}[h]
\begin{center}
\begin{tabular}{ |c|c|c|c|} \hline
 Observable   &General & Airy & Zeta \\ \hline
Master Matrix &  $M_0$ & $a + a^+$ & ?\\
Potential & $V(M)$ & $Tr(M^2)$ & $
\mathop {\lim }\limits_{p \to \infty } Tr(V_p (M) + \sum\limits_{k = 1}^{p - 2} {s_k V_k (M))}$
 \\
FZZT Brane & $B(z)$ & $Ai(z)$ & $\zeta( iz +1/2)$\\
Macroscopic Loop & $W(z)$ & $\log{Ai(z)}$ & $\log{\zeta(iz+1/2)}$ \\
Kontsevich Integrand & $e^{-U(\phi)}$ & $e^{i\phi^3/3}$ & 
$\sum\limits_{q = 1}^\infty(q^4 \pi ^2 e^{2\phi}  - \frac{3}{2} q^2 \pi e^\phi ) \exp(-\pi q^2 e^\phi)$\\ \hline
\end{tabular}
\end{center}
\caption{Analogy between the Airy function and the Riemann zeta function. The quantities $V_k(M)$ and $s_k$ defined by a generalized $(p,1)$ matrix model in the following section. }
\label{tab1}
\end{table}

The integral representation of the Airy function has a matrix integral generalization. 
The matrix potential is defined from:
\[
e^{ - U(\Phi )}  =  e^{i\frac{1}{3}Tr(\Phi ^3 )}
\]
The matrix generalized Airy function is given by:
\[
Ai(Z) = \int d\Phi e^{iTr(Z\Phi )}e^{-U(\Phi)}
\]
In the above $\Phi$ and $Z$ are $n \times n$ matrices. The interpretation of this matrix integral is that it describes $n$ FZZT branes.
The matrix $\Phi$ in the Kontsevich integrand is an effective degree of freedom describing open strings stretched between $n$ FZZT branes \cite{Kontsevich:1992ti}\cite{Kharchev:1992dj}\cite{Aganagic:2003qj}\cite{Gaiotto:2003yb}.
 
One can try to interpret the integrand of the $\Xi(z)$ function in a similar manner. In that case the analog of the potential defined by:
\[
e^{ - U(\Phi )}  =  \sum\limits_{k = 1}^\infty  {(\pi ^2 k^4 } e^{2Tr\Phi }  - \frac{3}{2}\pi k^2 e^{Tr\Phi } )e^{ - \pi k^2 Tre^\Phi  }
\]
and the analog of the the matrix integral describing $n$ FZZT branes is:
\[
\Xi[Z] = \int {D\Phi e^{iTr(Z\Phi )} } \sum\limits_{k = 1}^\infty  {(\pi ^2 k^4 }
e^{2Tr\Phi }  - \frac{3}{2}\pi k^2 e^{Tr\Phi } )e^{ - \pi k^2 Tre^\Phi  }
\]
This is the origin of the matrix model given by (3.1). As discussed in section 2 this can treated as a sum of Liouville type matrix models.

\section{Relation to generalized $(p,1)$ matrix models}

The Airy function is the FZZT partition function for the $(2,1)$ minimal matrix model.
In  \cite{Hashimoto:2005bf} the FZZT partition function was  given for the generalized $(p,1)$ minimal matrix model with parameters $s_k$. This theory has a characteristic polynomial or FZZT partition function given by:
\[
B(z) = \frac{1}{{2\pi }}\int {d\phi e^{iz\phi  - \frac{1}{{p + 1}}(i\phi )^{p + 1}  + \sum\nolimits_{k = 1}^{p - 2} {s_k \frac{1}{{k + 1}}(i\phi )^{k + 1} } } }
\]
Unlike the $(2,1)$ matrix model the definition of the generalized $(p,1)$ matrix model requires a two matrix integral of the form \cite{Hashimoto:2005bf}\cite{Daul:1993bg}\cite{Kazakov:2004du}\cite{Alexandrov:2003qk}\cite{Kharchev:1991cu}\cite{Alexandrov:2006sb}\cite{Mironov:2005qn}:
\[
Z_{(p,1)} (g) = \int {DMDAe^{ - \frac{1}{g}(V(M + I) - AM)} }
\]
Comparison with the integral representation of the $\Xi(z)$ function shows that a generalized matrix model for large $p$ can be constructed as an  approximation.
This can be compared with the formulas from the previous section to compute the corresponding coefficients $s_k$. One writes:
\[
\log \left( {\sum\limits_{k = 1}^\infty  {(\pi ^2 k^4 } e^{2\phi }  - \frac{3}{2}\pi k^2 e^\phi  )e^{ - \pi k^2 e^\phi  } } \right) =  - \frac{1}{{p + 1}}(i\phi )^{p + 1}  + \sum\nolimits_{k = 1}^{p - 2} {s_k \frac{1}{{k + 1}}(i\phi )^{k +
1} }
\]
In the above formula the function on the left is expanded to order $p+1$ in the variable $\phi$. We denote this terminated expansion by $\Xi_p(z)$.
Another way to compute the coefficients $s_k$is to differentiate the left hand side and set:
\[
s_k  = \frac{{i^{ - (k + 1)} }}{{k!}}\left. {\partial _\phi ^k \log \left( {\sum\limits_{k = 1}^\infty  {(\pi ^2 k^4 } e^{2\phi }  - \frac{3}{2}\pi k^2 e^\phi
)e^{ - \pi k^2 e^\phi  } } \right)} \right|_{\phi  = 0}
\]
From the integral representation one has:
\[
\begin{array}{l}
 Q\Xi _p (z) = z\Xi _p (z) \\
 P\Xi _p (z) =  - \partial _z \Xi _p (z) \\
 \end{array}
\]
where:
\[
Q = (P^p  + \sum\limits_{k = 0}^{p - 1} {s_k P^k } )
\]
Inserting this operator into the above equation one has the generalization of the Airy equation given by:
\begin{equation}
(P^p  + \sum\limits_{k = 0}^{p - 1} {s_k P^k } )\Xi _p (z) = z\Xi _p (z)
\end{equation}
To recover the equation for the full $\Xi(z)$ function one has to take $p$ to infinity which agrees with the fact that the zeta function does not obey a finite order differential equation. 

Note that $z$ and $\phi$ are in some sense canonically conjugate \cite{Hashimoto:2005bf}. Denote the Fourier transform of the $\Xi(z)$ function as $\tilde \Xi(p)$ then:
\[
\Xi (z) = \int {d\phi e^{i\phi z} \tilde \Xi } (\phi )
\]
The generalized Airy equation them becomes in Fourier space:
\[
(\phi ^p  + \sum\limits_{k = 0}^{p - 1} {s_k \phi ^k } )\tilde \Xi _p (\phi ) =
Q\tilde \Xi _p (\phi )
\]
This can be written:
\begin{equation}
(U'(\phi ) - Q)\tilde \Xi (\phi ) = 0
\end{equation}
where:
\[
e^{ - U(\phi )}  = \sum\limits_{k = 1}^\infty  {(\pi ^2 k^4 } e^{2\phi }  - \frac{3}{2}\pi k^2 e^\phi  )e^{ - \pi k^2 e^\phi  }
\]
Equation (5.2) is very similar to the equation for the master matrix (2.1). Indeed if we set:
\[
\begin{array}{l}
 \phi  = M_0 (y) \\
 z = P_0 (y) \\
 \end{array}
\]
we see that $y$ can be thought of as coordinates of a parametrization of the Riemann surface $M_{p,1}$ which is determined from the $\phi$ and $z$ constraint  $U'(\phi ) - z = 0$. If we make these variables into operators through:
\[
\begin{array}{l}
 \hat \phi  = \hat M_0 (a,a^ +  ) \\
 \hat z = \hat P_0 (a,a^ +  ) \\
 \end{array}
\]
this classical surface is turned into a quantum Riemann surface similar to those studied using noncommunative geometry \cite{Connes:1994yd}.

Once one has obtained the coefficients $s_k$ one can define matrix potential associated with a finite $N$ theory as \cite{Hashimoto:2005bf}:
\[
V(M) = \mathop {\lim }\limits_{p \to \infty } Tr(V_p (M) + \sum\limits_{k = 1}^{p - 2} {s_k V_k (M))}
\]
where:
\[
V_k (M) = \sum\limits_{j = 1}^p {\frac{1}{j}(M^j }  - I)
\]
This is the matrix potential of Table 1 in the previous section. 

A set of orthogonal polynomials with this matrix potential through the integral equation:
\[
B_n (z) = \frac{{n!}}{{2\pi i}}\oint {e^{ - \mathop {\lim }\limits_{p \to \infty } (V_p (y + 1) + \sum\limits_{k = 1}^{p - 2} {s_k V_k (y + 1))}  + 2zy} \frac{1}{{y^{n + 1} }}} dy
\]
Or equivalently though the generating function definition:
\[
e^{ - \mathop {\lim }\limits_{p \to \infty } (V_p (y + 1) + \sum\limits_{k =
1}^{p - 2} {s_k V_k (y + 1))}  + 2zy}  = \sum\limits_{n = 0}^\infty  {B_n (z)\frac{{y^n }}{{n!}}}
\]
These are the generalizations of the integral and generating function definitions of the Hermite polynomials associated with the $(2,1)$ minimal model.

We  note that some other entire functions can be treated in a similar manner. For example the reciprocal factorial function $\frac{1}{\Pi(z)} = \frac{1}{\Gamma(z+1)}$
has a product representation:
\[
\frac{1}{\Pi (z)} = e^{\gamma z} \prod\limits_{n = 1}^\infty  {(1 + \frac{z}{n}} )e^{ - z/n}
\]
and integral representation:
\[
\frac{1}{\Pi (z)} = \int {d\phi e^{iz\phi } e^{ - e^\phi  } }
\]
The identity 
\[
\frac{1}{\Pi (z - 1)} = z\frac{1}{{\Pi (z)}}
\]
implies the equation
\[
e^{ - \partial _z } \frac{1}{\Pi (z)} = z\frac{1}{{\Pi (z)}}
\]
or:
\[
(e^P  - z)\frac{1}{\Pi (z)} = 0
\]
This is the analog of equations (5.1) for the $\Xi(z)$ function. The product representation shows that the zeros of the inverse factorial function are of the form $\lambda _n  =  - 1, - 2, - 3, \ldots$. The inverse factorial function is similar to the zeta function in that it does not obey a finite order differential equation. It is similar to the Airy function in that it has all it's zeros on the negative real axis. It differs from both the Airy and zeta function in that it's zeros are of a simple form namely the negative integers.
The integral representation of the reciprocal factorial function seems related to the Liouville matrix model with Kontsevich integrand $e^{-U(\phi)} = e^{-e^\phi}$ and generalized $(p,1)$ matrix model with $s_k= \frac{1}{k!}$. The matrix integral representation of the Gamma function in terms of the Liouville matrix model has been discussed in \cite{Mukhi:2003sz}.

Most of our analysis has centered on the matrix side of the matrix/gravity correspondence. The gravity side is related through an integral transform. For example the macroscopic loop observable associated with the Riemann zeta function is given by:
\[
\log \zeta (iz + 1/2) = \int\limits_0^\infty  {\ell ^{ - iz - 1/2} W(\ell )d\ell }
\]
In terms of the $\lambda_n$ this observable takes the form \cite{Edwards}:
\[
W(\ell ) = \frac{1}{{\log \ell }} - \sum\limits_n {\frac{{2\cos (\lambda _n \log \ell )}}{{\ell ^{1/2} \log \ell }}}  - \frac{1}{{\ell (\ell ^2  - 1)\log \ell }}
\]
The indefinite integral of this Wheeler-DeWitt wave function is connected to the prime numbers $p$ through:
\[
\int\limits_2^\ell  {W(\ell ')d\ell '}  = \frac{1}{2}(\sum\limits_{p^n  < x} 
{\frac{1}{n}}  + \sum\limits_{p^n  \le \ell} {\frac{1}{n}} )
\]
The FZZT brane partition function can also be represented by prime numbers as:
\[
\log \zeta (iz + 1/2) = \sum\limits_p {\sum\limits_n {\frac{1}{n}p^{ - n(iz + 1/2)} } }
\]
Both of the above formulas follow from the Euler product formula of the zeta function. Much of the physical intuition about the meaning of the FZZT brane and the Wheeler-DeWitt wave function occurs on the gravity side of the correspondence. Thus the connection of number theory and gravity in this context is quite intriguing.

Finally to approach the generalized Riemann hypothesis using the \\
matrix/gravity correspondence one can replace the Kontsevich integrand
$e^{-U(\phi)}$ with a modular function. Indeed such modular functions already arise in the matrix/gravity $CFT_2/AdS_3$ correspondence between two dimensional conformal field theory and three dimensional gravity with negative cosmological constant \cite{Witten:2007kt}. 

\section{Conclusion}

In this paper we have examined the Riemann zeta function as a FZZT brane partition function involved in matrix models. The FZZT description gives rise to the physical interpretation of the Riemann hypothesis, that the $\Xi(z)$ is an entire function and has zeros on the critical line with $z$ on the real axis (this corresponds to the $Re(s) = \frac{1}{2}$). The zeros are interpreted as eigenvalues of the master matrix. The macroscopic loop observable and resolvent also have physical interpretations in terms of the matrix model. In the gauge gravity correspondence the macroscopic loop is identified with the Wheeler-DeWitt wave function of the 2d world sheet gravity. The variable $z$ is identified with the boundary cosmological constant in the 2d gravity. The matrix gravity correspondence is the mapping between the matrix quantities and the 2d gravity computations. In a string theory context these in turn describe target space time processes.  

The Kontsevich integrand was identified using the Fourier transform of the $\Xi(z)$ function. Replacing the $z$ variable by $n\times n$ matrix $Z$ and the Kontsevich integrand by a matrix integrand one obtains representation of a matrix model describing $n$ FZZT branes. The Kontsevich integrand is given by a superposition Liouville matrix models that have been used to represent instanton matrix models for the $c=1$ string. 

Some  long standing  issues are indicated in Table 1. To identify and interpret the Master matrix associated with the Riemann zeta function.

\end{document}